\documentclass[final,5p,times,twocolumn]{elsarticle}
\pdfoutput=1
\usepackage{preprintcover}
\usepackage{graphicx}
\usepackage{lineno}
\usepackage{url}
\usepackage{hyperref}
\usepackage{mathptmx}
\usepackage{amssymb}
\usepackage{amsmath}
\usepackage{verbatim}
\usepackage{subfigure}
\usepackage{placeins}
\usepackage{xspace}
\usepackage[labelfont=bf,tableposition=top]{caption}
\newcommand{\vpion}{\ensuremath{\pi_{\rm v}}\xspace}
\newcommand{\vpions}{\vpion\xspace}
\newcommand{\mvpion}{\ensuremath{m_{\vpion}}\xspace}
\newcommand{\sigmaSM}{\ensuremath{\sigma_{\rm SM}}\xspace}

\newcommand{\higgsd}{\mbox{$H\rightarrow\vpion\vpion$}\xspace}

\newcommand{\intlumi}{20.3~fb$^{-1}$\xspace}

\newcommand{\rts}{\ensuremath{\sqrt{s}}}
\newcommand{\pp}{\ensuremath{pp}\xspace}
\newcommand{\et}{\ensuremath{E_{\rm{T}}}\xspace}
\newcommand{\pt}{\ensuremath{p_{\textrm{T}}}\xspace}

\newcommand{\met}{\ensuremath{E_{\mathrm{T}}^{\mathrm{miss}}}\xspace}
\newcommand{\calrat}{\ensuremath{{{ \log}_{10}(E_{\rm{\rm{H}}}/E_{\rm{EM}})}}\xspace}

\newcommand{\eh}{\ensuremath{E_{\rm{H}}}\xspace}
\newcommand{\eem}{\ensuremath{E_{\rm{EM}}}\xspace}
\newcommand{\timingcut}{$-1<t< 5$~ns\xspace}
\newcommand{\ntrack}{\ensuremath{n_{\rm track}}\xspace}
\newcommand{\deltaphi}{\ensuremath{\Delta\phi}\xspace}

\newcommand{\PhiHS}{\ensuremath{\Phi_{\rm{hs}}}\xspace}

\newcommand{\PhiHSd}{\mbox{$\PhiHS\rightarrow\vpion\vpion$}\xspace}
\newcommand{\Phid}{\mbox{$\Phi\rightarrow\vpion\vpion$}\xspace}

\newcommand{\PhiSM}{\ensuremath{\Phi}\xspace}
\newcommand{\mPhiSM}{\ensuremath{m_{\Phi}}\xspace}
\newcommand{\mHiggs}{\ensuremath{m_{{H}}}\xspace}

\newcommand*{\TeV}{\ifmmode {\mathrm{\ Te\kern -0.1em V}}\else
                   \textrm{Te\kern -0.1em V}\fi\xspace}
\newcommand*{\GeV}{\ifmmode {\mathrm{\ Ge\kern -0.1em V}}\else
                   \textrm{Ge\kern -0.1em V}\fi\xspace}
\newcommand*{\MeV}{\ifmmode {\mathrm{\ Me\kern -0.1em V}}\else
                   \textrm{Me\kern -0.1em V}\fi\xspace}
\newcommand*{\keV}{\ifmmode {\mathrm{\ ke\kern -0.1em V}}\else
                   \textrm{ke\kern -0.1em V}\fi\xspace}
\newcommand*{\eV}{\ifmmode  {\mathrm{\ e\kern -0.1em V}}\else
                   \textrm{e\kern -0.1em V}\fi\xspace}
\let\tev=\TeV
\let\gev=\GeV

\newcommand{\titletext}{Search for pair-produced long-lived neutral particles decaying to jets in the ATLAS hadronic calorimeter in $pp$ collisions at $\rts= 8$~TeV}
\PreprintCoverPaperTitle{\titletext}
\PreprintIdNumber{CERN-PH-EP-2014-228}
\newcommand{\abstxt}{The ATLAS detector at the Large Hadron Collider at CERN is used to search for the decay of a scalar boson to a pair of long-lived particles, neutral under the Standard Model gauge group, in \intlumi of data collected in proton--proton collisions at $\rts = 8$~\tev. This search is sensitive to long-lived particles that decay to Standard Model particles producing jets at the outer edge of the ATLAS electromagnetic calorimeter or inside the hadronic calorimeter. No significant excess of events is observed. Limits are reported on the product of the scalar boson production cross section times branching ratio into long-lived neutral particles as a function of the proper lifetime of the particles. Limits are reported for boson masses from 100 \GeV to 900 \GeV, and a long-lived neutral particle mass from 10 \GeV to 150 \GeV.}
\PreprintCoverAbstract{\abstxt}
\PreprintJournalName{Physics Letters B}

\def\@atlasnote{}
\newcommand{\atlasnote}[1]{\def\@atlasnote{#1}}

\ifx\@atlasnote\@empty \vskip 2mm \mbox{ }
\else \vskip 4mm {\Large \@atlasnote} \vskip 2mm
\fi

\atlasnote{ATLAS-CONF-2014-058}

\begin{document}

\biboptions{sort&compress}

\begin{frontmatter}

\title{\titletext}

\author{The ATLAS Collaboration}

\begin{abstract}
\abstxt
\end{abstract}

\end{frontmatter}

\section{Introduction}\label{introduction}

The discovery of the Higgs boson~\cite{HIGGS1,HIGGS2,HIGGS3} by the ATLAS and CMS experiments~\cite{ATLASdiscovery,CMSdiscovery} in 2012 identified the last piece of the highly successful Standard Model (SM). Subsequent measurements of the Higgs boson branching ratios and couplings, while consistent with the SM expectations, allow for a substantial branching ratio to exotic particles. This letter describes a search for decays of the Higgs boson and other scalar bosons to non-SM states that in turn decay to SM particles.

A number of extensions of the SM involve a hidden sector that is weakly coupled to the SM, with the two connected via a communicator particle.  This letter considers models containing a hidden sector with a confining gauge interaction that is otherwise invisible to the SM. The communicator is chosen to be a SM-sector scalar boson, \PhiSM~\cite{Chung2006,Chang2008,HV1,HV2}. The communicator mixes with a hidden-sector scalar boson, \PhiHS, which decays into detectable SM particles. This search considers communicator masses between 100~\gev and 900~\gev. A \PhiSM mass close to the mass of the discovered Higgs boson is included to search for exotic decays of the Higgs boson‏.

A Hidden Valley (HV) model~\cite{HV1,HV2} is used as the benchmark model. The lightest HV particles form an isospin triplet of pseudoscalar particles which are called valley pions (\vpion) because of their similarity to the SM triplet. The \vpion are pair-produced ($\PhiHSd$) and each decays to a pair of SM fermions. The \vpion possess Yukawa couplings to fermions and therefore preferentially decay to accessible heavy fermions, primarily $b\overline{b}$, $c\overline{c}$ and $\tau^{+}\tau^{-}$.

The lifetime of the \vpion~is unconstrained and could be quite long. A long-lived \vpion can result in signatures that traditional searches fail to detect.  If a \vpion decays in the inner detector or muon spectrometer, it can be reconstructed as a displaced vertex. However, standard vertex-finding algorithms~\cite{chargedVertexing} are not likely to reconstruct it without modification. Likewise, a \vpion decay deep inside the calorimeter is reconstructed as a jet with an unusual energy signature that most traditional searches reject as having poor data quality. This search focusses on final states where both \vpion~decay in the hadronic calorimeter or near the outer edge of the electromagnetic calorimeter. Each heavy fermion pair from a \vpion decay is reconstructed as a single calorimeter jet with three characteristic properties: a narrow radius, no tracks from charged particles matched to the jet, and little or no energy deposited in the electromagnetic calorimeter.

Scalar boson masses ranging from 100~\gev to 900~\gev are considered in addition to the Higgs boson's mass (generated at $\mHiggs=126$~GeV) and \vpion~masses between 10~\gev and 150~\gev are studied. Other searches for pairs of displaced vertices generated by pair-produced neutral, long-lived particles were performed in‏‏ ATLAS~\cite{dan} and CMS~\cite{moriond} at the LHC and in D0~\cite{HVD0} and CDF~\cite{HVCDF} at the Tevatron. The Tevatron experiments and CMS searched for displaced vertices in their tracking system only, which results in a corresponding proper decay length range of a few meters. CMS also looked at the multi-lepton decay channel, another possible decay of HV particles. The previous ATLAS analysis, based on 7~\tev data, used the muon spectrometer and is sensitive to proper decay lengths between 0.5~m and 27~m, depending on the benchmark model. No evidence of physics beyond the SM was found.

\section{The ATLAS detector}

The ATLAS detector~\cite{ATLASTDR} is a multi-purpose detector at the
LHC, consisting of several sub-detectors. From
the interaction point (IP) outwards there are an inner detector (ID), electromagnetic and
hadronic calorimeters, and a muon spectrometer (MS).
The ID, immersed in a 2 T axial magnetic field, provides tracking and vertex information for charged
particles within the pseudorapidity \footnote{ATLAS uses a right-handed coordinate system with its origin at the nominal interaction point (IP) in the centre of the detector and the $z$-axis along the beam pipe. The $x$-axis points from the IP to the centre of the LHC ring, and the $y$-axis points upward. Cylindrical coordinates $(r,\phi)$ are used in the transverse plane, $\phi$ being the azimuthal angle around the beam pipe. The pseudorapidity is defined in terms of the polar angle $\theta$ as $\eta=-\ln\tan(\theta/2)$.} ($\eta$) region $|\eta| <$ 2.5.
It consists of three different tracking detectors. From small radii outwards, these are
a silicon pixel detector, a silicon microstrip tracker (SCT) and a transition radiation tracker (TRT).

 The calorimeter provides coverage over the range $|\eta| < 4.9$. It consists of a lead/liquid-argon electromagnetic calorimeter (ECal) at smaller radii surrounded by a hadronic calorimeter (HCal) at larger radii comprising a steel and scintillator-tile system in the barrel region ($|\eta| < 1.7$) and a liquid-argon system with copper absorbers in the endcaps ($1.5 < |\eta| < 3.2$). The ECal spans the range $1.5 {\rm ~m} < r < 2.0$~m in the barrel and $3.6 {\rm ~m} < |z| < 4.25$~m in the endcaps. The HCal covers $2.25 {\rm ~m} < r < 4.25$~m in the barrel and $4.3{\rm ~m} < |z| < 6.05$~m in the endcaps. There is also a forward calorimeter (FCal), with coverage between $3.1 < |\eta| < 4.9$, which uses copper absorbers in the first layer, and tungsten absorbers in the second and third layers, and liquid-argon as the active medium in all layers.
Muon identification and momentum measurement are provided
by the MS, which extends to $|\eta| =$ 2.7. It consists of a three-layer system of gas-filled precision-tracking
chambers. The region  $|\eta| <$ 2.4 is also covered
by separate trigger chambers.

A sequential three-level trigger system selects events to be recorded for offline analysis. The first level consists of custom hardware that implements selection on jets, electrons, photons, $\tau$ leptons, muons, and missing transverse momentum or large total transverse energy. The second and third levels add charged particle track finding and refine the first-level selections with progressively more detailed algorithms.

\section{Data and simulation samples}

All data used in this analysis were collected during the 2012 LHC proton-proton run at a centre-of-mass energy of  8~\tev. After data quality requirements are applied, the sample corresponds to an integrated luminosity of \intlumi.
The HV Monte Carlo (MC) samples are generated with PYTHIA 8.165~\cite{PYTHIA} and the PDF MSTW2008‏~\cite{mstw} to simulate gluon fusion $gg\to \PhiSM$ production and the \PhiHS decay \PhiHSd for different \PhiSM and \vpion~masses (Table~\ref{tab:mcparams}). \PhiSM masses below 300~\GeV are considered low-mass samples and the rest are considered high-mass samples.  The \vpion lifetime is fixed in each sample to ensure decays throughout the ATLAS detector.
The \PhiSM is simulated in PYTHIA by replacing the Higgs boson with the \PhiSM and having the \PhiSM decay to \vpion 100\% of the time.
The \PhiSM samples are produced with cross sections calculated at next-to-next-to-leading-logarithmic accuracy in QCD processes and at next-to-leading-order in electro-weak processes assuming the \PhiSM at each mass has the same properties as the SM Higgs boson~\cite{HiggsXs}.
After generation the events are passed through a detailed simulation of the detector response with GEANT4~\cite{ATLASG4,Agostinelli:2002hh} and the same reconstruction algorithms as are used on the data. GEANT4 needed no modification to simulate the signal as all decay particles are SM particles.
All MC samples are reweighted to reproduce the number of interactions per bunch crossing observed in the data.

\begin{table}[ht!]
  \caption{The \PhiSM mass or Higgs boson mass, \PhiSM gluon fusion production cross section, and \vpion mass of each benchmark Hidden Valley model generated. The cross-sections are based on the assumption in the benchmarked model that the $\Phi$ boson production mechanism is the same as the Higgs boson production mechanism. The decay branching ratios of the \vpions as a function of the \vpion mass are listed in the second table as determined in the simulation samples.}
  \label{tab:mcparams}
  \centering
\subfigure{
\begin{tabular}{c|c|c}
\hline

\mHiggs & $\sigma$ & \vpion~Mass \\
{[}\GeV] & [pb] & [\GeV] \\ \hline \hline
 126 & 19.0 & 10, 25, 40  \\  \hline \hline

\end{tabular}
}

\subfigure{
\begin{tabular}{c|c|c}
\hline

\PhiSM Mass & $\sigma$ & \vpion~Mass \\
{[}\GeV] & [pb] & [\GeV] \\ \hline \hline
 100 & 29.7 & 10, 25 \\
 140 & 15.4 & 10, 20, 40  \\
 300 & 3.59 & 50 \\
 600 & 0.52 & 50, 150 \\
 900 & 0.06 & 50, 150 \\ \hline \hline
\end{tabular}
}
\subfigure{
\begin{tabular}{c|c|c|c}
\hline
\vpion~Mass & BR $b\overline{b}$ & BR $\tau^{+}\tau^{-}$ & BR $c\overline{c}$\\
 {[}\GeV] & [\%] & [\%] & [\%]\\\hline\hline
 10  & 70.0 & 16.4 & 13.4\\
 20  & 86.3 &  8.0 &  5.6\\
 25  & 86.6 &  8.1 &  5.3\\
 40  & 86.5 &  8.5 &  5.0\\
 50  & 86.2 & 8.8 & 4.9\\
 150  & 84.8 & 10.2 & 4.8\\\hline\hline
\end{tabular}
}
\end{table}

\section{Trigger and event selection}

Candidate events are collected using a dedicated trigger, called the {\em CalRatio} trigger~\cite{trigPaper2012}, which looks specifically for long-lived neutral particles that decay near the outer radius of the ECal or within the HCal. The trigger is tuned to look for events containing at least one narrow jet with little energy deposited in the ECal and no charged tracks pointing towards the jet. At the first level the trigger selects only narrow jets by requiring at least 40~\gev of transverse energy ($\et$) in the calorimeter in a $0.2\times 0.2$ ($\Delta\eta\times\Delta\phi$) region using topological jets~\cite{ATLASTDR,L1CAL}, in contrast to the default algorithm in which the energy in a $0.4\times 0.4$ region is summed.
The 40 \GeV \et threshold requirement is fully efficient at an offline jet \et of 60~\GeV.
 To select jets with a high fraction of their energy in the HCal the second level of the trigger requires these narrow jets to have $\calrat>1.2$, where $\eh/\eem$ is the ratio of the energy deposited in the HCal (\eh) to the energy deposited in the ECal (\eem).
The trigger also requires no tracks with $p_{\rm{T}}>1$~\gev in the region $0.2\times 0.2$ ($\Delta\eta \times \Delta\phi$) around the jet axis. The third level of the trigger uses the slower but more accurate anti-$k_t$ algorithm~\cite{jetalgo} with $R=0.4$ to reconstruct the jet and requires the jet to have a minimum of 35~\gev of transverse energy.

\begin{figure}[!ht]
\subfigure[]
{\includegraphics[width=0.48\textwidth]{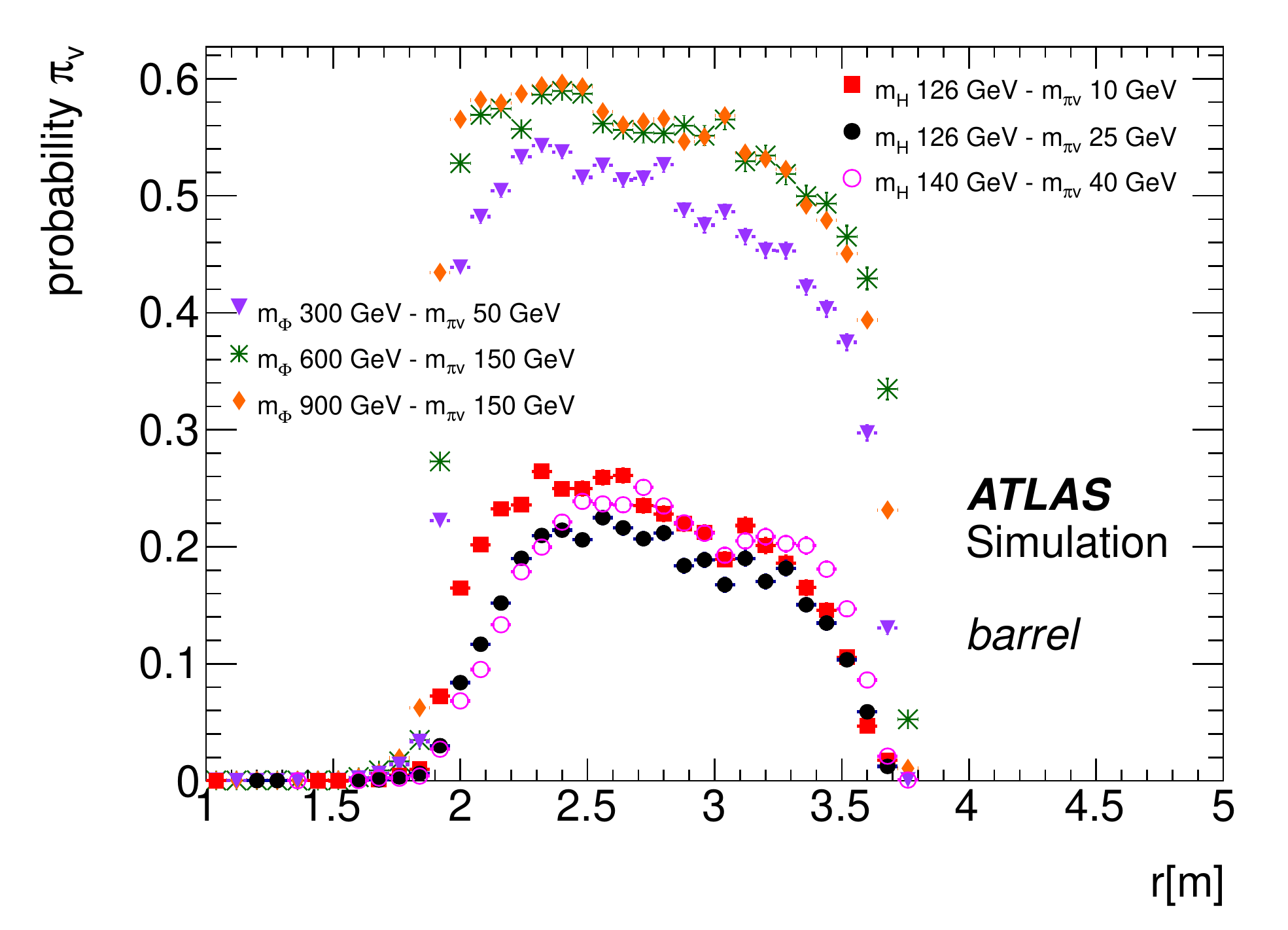}}
\subfigure[]
{\includegraphics[width=0.48\textwidth]{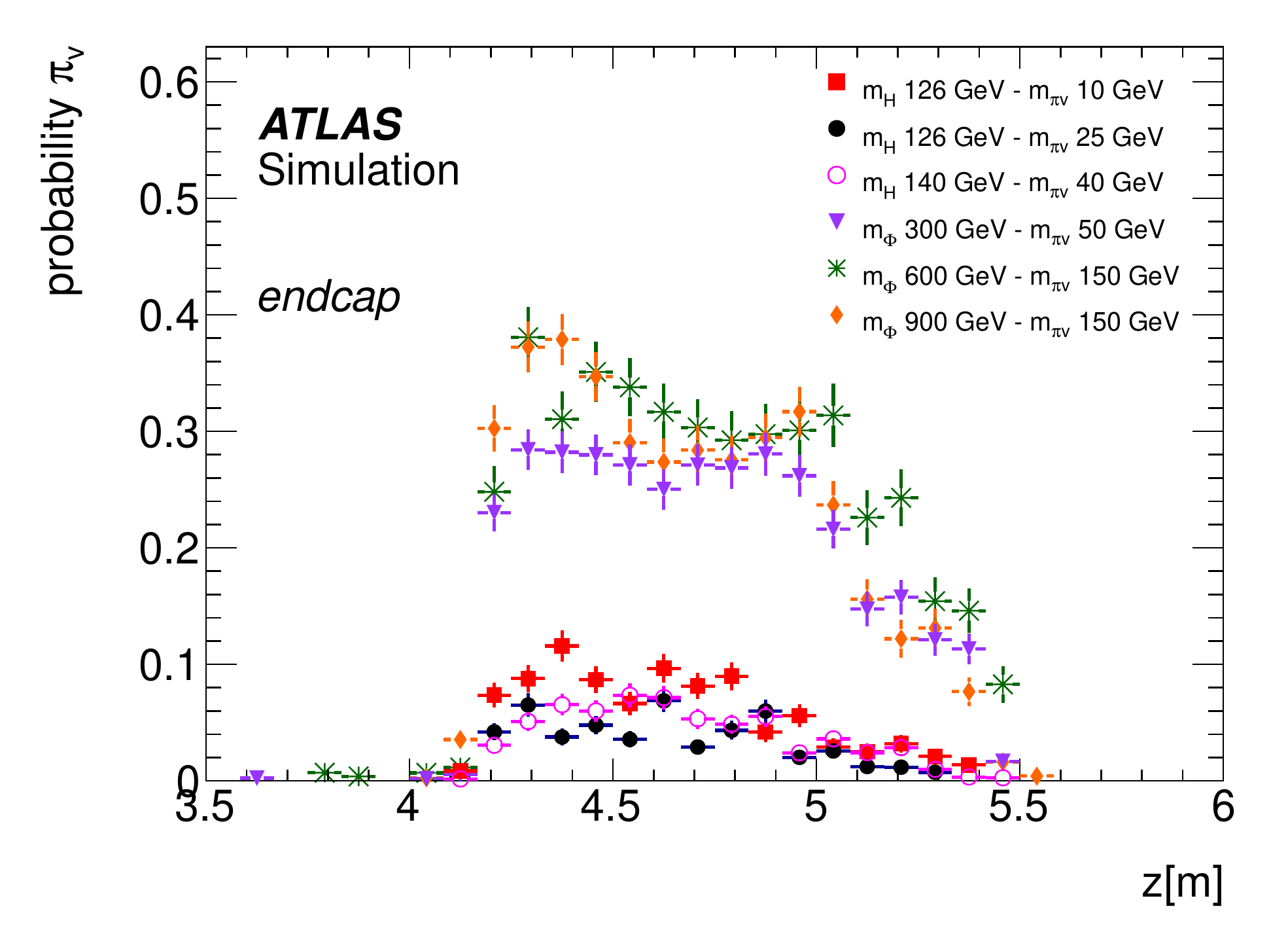}}
\caption{The probability ($\varepsilon_{\pi_{\rm v}}$) for a single \vpion to pass the trigger as a function of the \vpion (a) radial decay length in the barrel and (b) the $z$ position of the decay vertex in the endcaps for several \PhiSM and \vpion masses.
}
\label{fig:trigEffi}
\end{figure}

The probability ($\varepsilon_{\pi_{\rm v}}$) for a single \vpion to fire the trigger in simulated events is shown in Fig.~\ref{fig:trigEffi}, for the (a) barrel and (b) endcap region of the calorimeter in several different signal samples. The average probability for the
low (high) scalar boson masses is about 20\% (55\%) for \vpion decays occurring at radii between 2.0~m and 3.5~m in the barrel, and about 6\% (30\%) for \vpion decays with $|z|$ between 4.0~m and 5.5~m in the endcaps.
The turn-on takes place before the inner edge of the HCAL as the \calrat cut allows for a small amount of energy in the ECal. The probability decreases towards the outer region of the HCal where too much of the energy escapes the HCal to pass the jet \et requirement. The efficiency is lower in the endcaps because events tend to not satisfy the isolation criteria due to the increased occupancy from extra collision events in the same bunch crossing as a hard-scatter interaction (pile-up).

Events also contain a reconstructed primary vertex with at least three tracks with $\pt>1$~\GeV. Events are rejected if any reconstructed jets show evidence of being caused by a beam-halo interaction~\cite{trigPaper2012}. A missing transverse momentum requirement, \met $<$ 50~\gev, is applied to reject non-collision events, such as cosmic rays or beam-halo interactions.

In the offline selection, jets are reconstructed with an anti-$k_t$ algorithm with $R=0.4$, starting from calorimeter energy clusters calibrated using the local cluster weighting method~\cite{JES}. Jets are then calibrated using an energy- and $\eta$-dependent simulation-based calibration scheme.
Jets are rejected if they do not satisfy the standard ATLAS good-jet criteria with the exception of requirements that reject jets with small electromagnetic energy fraction (EMF)~\cite{JetClean}. At least one jet must have fired the CalRatio trigger.  The jet matching the trigger must pass an $\et > 60$~\GeV requirement while a second jet must satisfy an $\et > 40$~\GeV requirement.  If more than one jet fired the CalRatio trigger then only the leading jet is required to have $\et > 60$~\GeV.

\begin{figure}[!ht]
\subfigure[]
{\includegraphics[width=0.45\textwidth]{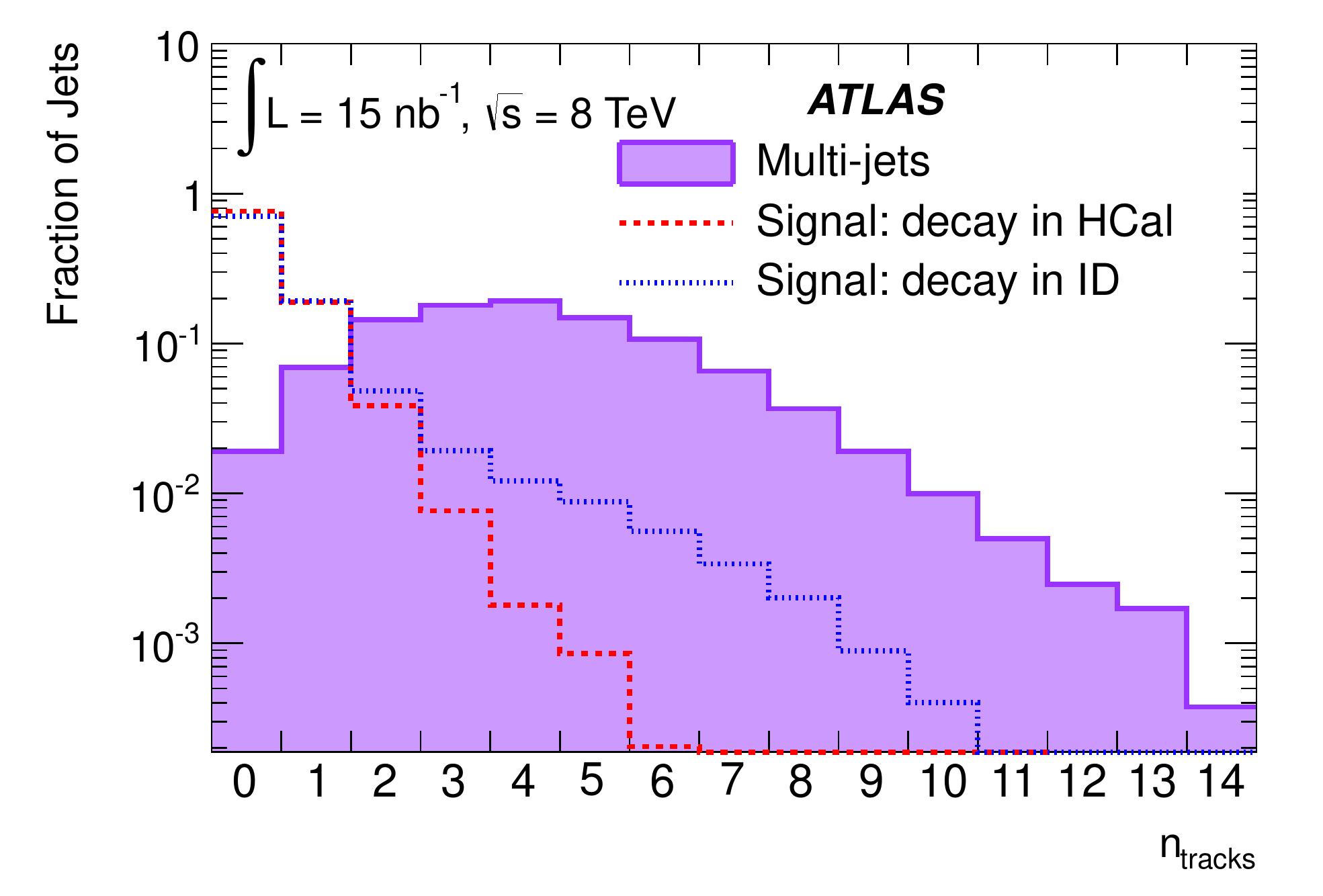}}
\subfigure[]
{\includegraphics[width=0.45\textwidth]{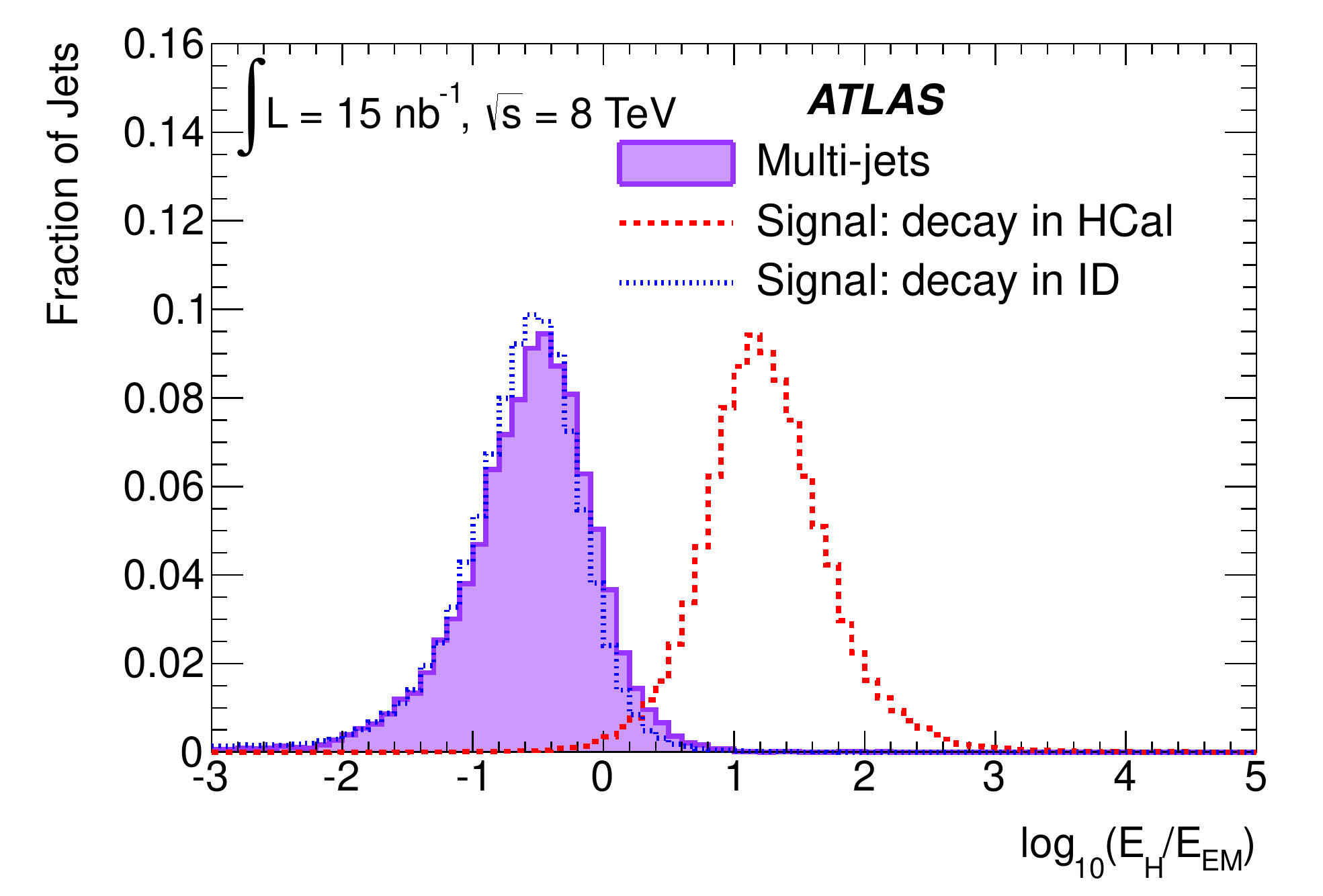}}
\caption{
Distribution of (a) the number of good tracks ($n_{\rm{tracks}}$) with $p_{\rm{T}} > 1$ \GeV and $\Delta R < 0.2$ around the jet axis
and (b) the distribution of jet \calrat with jet $|\eta| < 2.5$, $p_{\rm{T}} > 40$~\GeV.
The dashed histogram is for \vpion jets decaying in the hadronic calorimeter, and the dotted histogram is for \vpion jets decaying in the ID. Both are from the $\mHiggs = 126$~\GeV, $\mvpion = 10$~\GeV sample.
The filled histogram is the multi-jet data sample used to evaluate the multi-jet contribution to the background.
Events are required to satisfy $\met < 50$~\GeV.
}
\label{fig:QCD_HV}
\end{figure}

Individually, all jets must satisfy $|\eta| < 2.5$, have $\calrat>1.2$, and have no good tracks in the ID with \pt$>$1 \GeV in a region $\Delta R < 0.2$\footnote{$\Delta R = \sqrt{(\Delta\eta)^2 +(\Delta\phi)^2}$} centred on the jet axis. A good track must have at least two hits in the pixel detector and a total of at least nine hits in the pixel and SCT detectors.
Fig.~\ref{fig:QCD_HV}(a) compares the distribution of the number of good tracks associated with each jet in the multi-jet sample (described in the next section) with that in jets resulting from simulated $\pi_{\rm{v}}$ decays in the HCAL or ID. Fig.~\ref{fig:QCD_HV}(b) makes the same comparison for the distribution of $\calrat$ of each jet.
The multi-jet data was gathered using a prescaled, single-jet trigger with a 15 \GeV requirement.

Jets caused by cosmic rays and beam-halo interactions are often out-of-time. The jet timing is calculated by making an energy-weighted average of the timing for each cell in the jet. Each cell is defined to have a time of 0 ns if its energy is recorded at a time consistent with the arrival of a $\beta = 1$ particle from the IP. The timing of each jet is required to satisfy \timingcut. This cut will impact the efficiency for low $\beta$ \vpion. Due to the requirement of a high-\et jet in this analysis the $\beta$-distribution is peaked near 1 for low mass \PhiSM samples. For the high mass \PhiSM samples the difference between \mPhiSM and \mvpion results in a large boost for the \vpion at the generated lifetimes. As a result, the inefficiency introduced by the timing cut is at worst 1.5\% for the considered samples.

The analysis requires that exactly two jets satisfy these requirements. The second jet requirement significantly reduces the SM multi-jet background contribution. Table~\ref{tab:summary} lists the final number of expected events in each signal MC sample. The final number of events selected in data is 24.

\section{Background estimation }

The largest contribution to the expected background comes from SM multi-jet events. Cosmic-ray interactions contribute at a much lower level, and beam-halo interactions make a negligible contribution.

To estimate the multi-jet background contribution, a multi-jet data sample is used to derive the probability that a jet passes the trigger and analysis selection.  To obtain a raw background prediction, these jet probabilities are applied to a data sample that represents the multi-jet background before application of jet-level analysis selection. A correction to account for two-jet correlations is applied to this raw prediction to yield the final multi-jet background estimate.

The multi-jet data sample contains events that pass single-jet triggers with an $\et$ threshold of 15~\GeV or higher. These triggers were prescaled in 2012 and their effective luminosities range between 14.8~nb$^{-1}$and 454.1~nb$^{-1}$. The dataset is representative of the full \intlumi of data collected in 2012 and contains events from throughout the data collection period.  The events are required to pass the analysis $\met$ requirement and have at least two back-to-back ($\deltaphi > 2.0$) jets with $\et>40$~\GeV and $-2.5 < \eta < 2.5$. One jet is required to satisfy the modified ATLAS good-jet criteria used by the analysis. The second is used to measure two probabilities: one, called $P$, for a jet to pass the trigger and the $\et>60$~\GeV jet requirement and the other, called $Q$, for a jet to satisfy the requirement $\et>40$~\GeV. For both $P$ and $Q$ the jet must also pass the \calrat, track isolation, and all other analysis jet selection requirements including the modified ATLAS good-jet criteria. The probabilities are determined as a function of jet $\et$ and $\eta$.
The jet $\et$ and $\eta$ dependence is calculated independently because the sample is not large. A systematic uncertainty to account for any potential correlation is included in the analysis.
To calculate this systematic uncertainty the change in the mean \et as a function of $\eta$ and the change in the mean $\eta$ as a function of \et is measured in the multi-jet data sample. The maximum variation is 2\% for both $\et$ and $\eta$. The $\et$ or $\eta$ of each jet are systematically shifted by this amount as $P$ and $Q$ are recalculated. The new $P$ and $Q$ distributions are used to estimate the multi-jet background as described below, and the maximum variation in the result (6\%) is used as the systematic uncertainty.

\begin{figure}[!ht]
  \centering
  \subfigure[]
  {\includegraphics[width=0.48\textwidth]{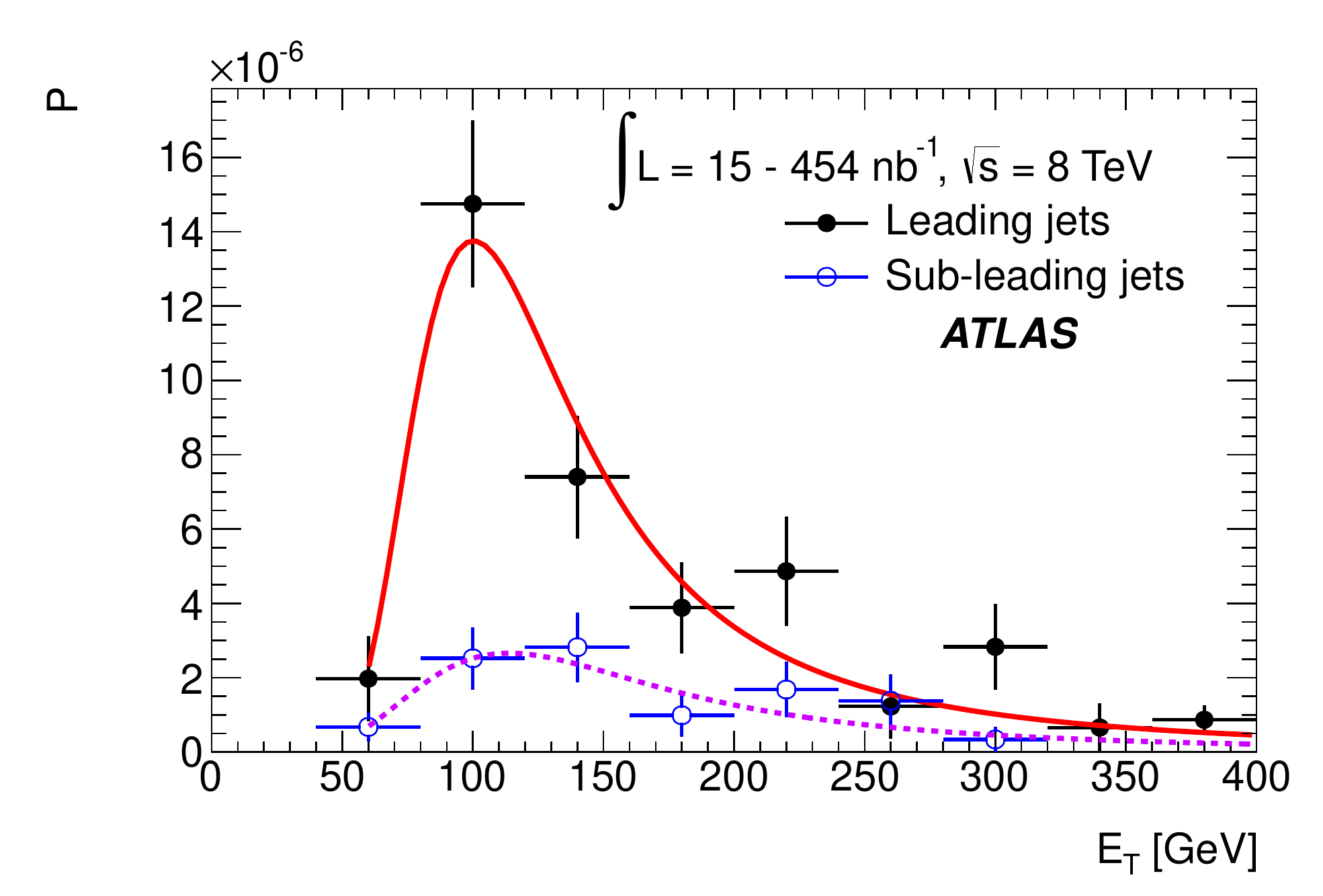}}
  \subfigure[]
  {\includegraphics[width=0.48\textwidth]{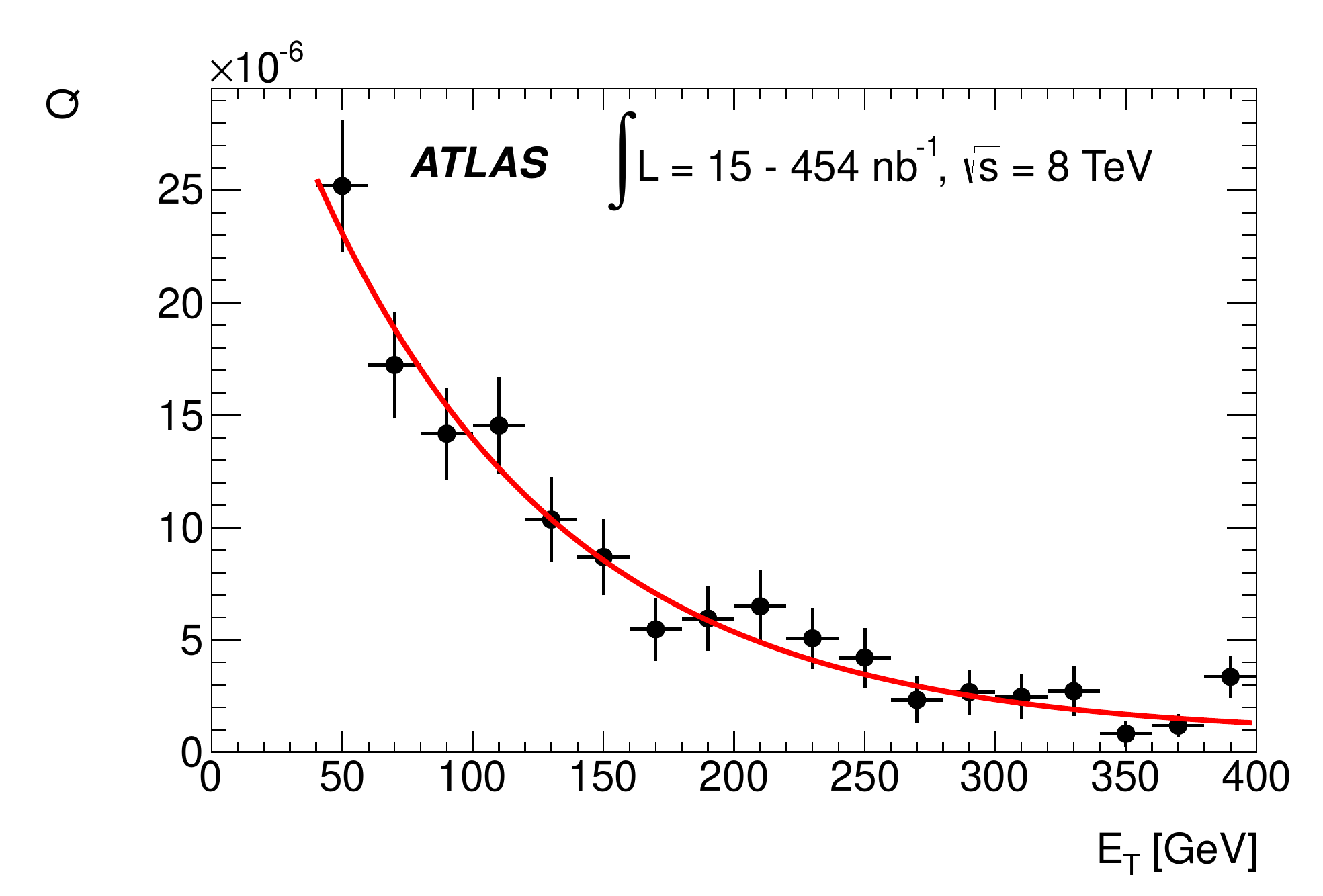}}
  \caption{The probability, $P$, that a jet from the multi-jet data sample passes the trigger and all jet requirements including the $\et>60$~\GeV requirement is shown in (a). A Landau function is fitted to the leading and sub-leading jet distributions separately (solid and dashed lines).
The probability, $Q$, to pass all jet requirements including the $\et>40$~\GeV requirement as a function of jet \et is shown in (b). An exponential function is fitted to the distribution (solid line). The $\et$ requirement is ignored when fitting to allow the curve to best match the shape of the data, but is used explicitly when $P$ and $Q$ are applied to a jet. The multi-jet data was gathered using a range of prescaled, single-jet triggers.}
\label{fig:multi-jetFit}
\end{figure}

Binned fits of the probabilities as a function of \et are made to a Landau function and an exponential function for $P$ and $Q$, respectively.
The $\et$ requirement is ignored when fitting to allow the curve to best match the distribution's shape.
The fit errors are propagated through to the systematic in the multi-jet background. The $\eta$ dependence is strongly correlated with the distribution of material in the calorimeters and cannot be well described by any simple functional form. Thus the probability is obtained directly from the distribution. The $P(\et)$ parameterisation is additionally split into leading jet and sub-leading jet samples because the probability is different for the two types of jets. This effect is also present for Q; however, it is accounted for by the correction for jet correlations discussed below. Plots of $P(\et)$ and $Q(\et)$ are shown in Fig.~\ref{fig:multi-jetFit}. The peak present in $P(\et)$ is the result of the trigger turn-on for the full trigger chain. The trigger jets are dominated by leading jets, and so dominated by the leading jet $P$.

The probability $P$ is verified using the CalRatio-triggered data. The CalRatio-triggered events are required to pass the same event selection used to derive the single-jet probabilities as well as the requirements for calculating $P$. The CalRatio-triggered data contains $501\,387$ events that fired the unprescaled CalRatio trigger and passed the required selection, and the single-jet probabilities predict $513\,000 \pm 94\,000$ (statistical error only) events.

To calculate the raw multi-jet background prediction the probabilities $P$ and $Q$ are applied to jets in events selected by the 15~\GeV single-jet trigger.
These single-jet probabilities are combined into an event probability using a combinatoric calculation that requires at least one jet in the event to fire the trigger and exactly two jets to pass all the jet selection criteria. The event probability is scaled to account for single-jet trigger prescales, yielding a weight for each event.  The sum of all weights in the data sample yields a raw background prediction of 13.2 $\pm$ 2.9 (statistical $\oplus$ systematic error) events. The uncertainty is dominated by the small number of jets firing the CalRatio trigger in multi-jet events.

In multi-jet events the \calrat and track isolation values of one jet are correlated with those of the second jet. If an event contains one jet of high \calrat, the second jet is more likely to have high \calrat as well.  Likewise, if one jet has no tracks associated with it, the other is more likely to have no associated tracks as well. The single-jet probabilities above ignore this correlation because each is calculated independently of the \calrat and \ntrack of other jets in the event. As a result, $Q$ is lower than if it were calculated only in events with an accompanying low \ntrack, high \calrat jet.

A scale factor to account for the correlation is calculated from the multi-jet data sample and the CalRatio-triggered data sample by examining numbers of events in regions in the \calrat and number-of-tracks ($\ntrack$) plane that are outside the signal region ($\calrat > 1.2$ and $\ntrack = 0$). The \calrat binning is chosen such that binning is uniform in EMF,\footnote{ $\calrat = {\rm \log}_{10}((1-{\rm EMF})/{\rm EMF})$.}. A range from 0 to 7 was used for \ntrack. The regions outside the signal region are expected to have very little signal contamination.

In each region the ratio of the number of events observed in the CalRatio-triggered data to the raw prediction is calculated.  Two series of ratios are calculated, one as a function of $\calrat$ and one as a function of $\ntrack$.  To determine the trend in the ratio as a function of $\calrat$ the \ntrack requirement is held constant: a jet is required to have 5 or 6 tracks.  The ratio is then determined for several non-overlapping ranges of $\calrat$.  The same procedure is used for \ntrack by requiring jets to have $0.55<{\rm EMF}<0.65$. Because the ratio is taken with respect to the observed data, this ratio will correct for any normalisation errors in $P$ or $Q$.

Both sets of ratios are fitted to allow extrapolation into the signal region.  The product of the two ratios in the signal region yields a scale factor to correct for the correlation between jets.  A systematic error is added to account for the assumption that the two ratios are uncorrelated.  The calculated scale factor is 1.8 $\pm$ 0.5. The uncertainty on the scale factor is due to the limited sample size.

To verify the procedure eight other bins on the \calrat and \ntrack plane were chosen and the full background prediction method was applied. Because signal contamination is negligible outside of the signal region, the predicted number of events can be directly compared to the number of events in the same \calrat--\ntrack region in the CalRatio-triggered data. In all cases the prediction is consistent with data to within one standard deviation.

The final multi-jet prediction is 23.2 $\pm$ 8.0 (statistical $\oplus$ systematic error) events in the signal region. The uncertainty is dominated by the statistical uncertainty, which is in turn dominated by the small number of jets matching the CalRatio trigger in the multi-jet data sample. The systematic contribution comes from the correlation between $\et$ and $\eta$ as well as from the inclusion of a requirement on $\Delta\phi$ (not used in the signal selection) in the determination of $P$ and $Q$.

Particles from a cosmic-ray shower may pass through and deposit energy in the calorimeter without passing through the ID.
These energy deposits can be reconstructed as trackless jets. The overall contribution to the expected background is reduced by the jet-timing and \met requirements.

The cosmic-ray background was studied using a trigger similar to the CalRatio trigger, but active only during an empty crossing. Each proton beam is divided up into buckets, most of which are filled with protons. An empty crossing occurs when an empty bucket in each beam coincides in the centre of the detector, and five buckets on either side in each beam are also empty.
Data gathered from these empty crossings are used to study backgrounds that are not beam related.

The analysis selection, with the exception of the jet-timing requirement and the good-vertex requirement, are applied to all events triggered in empty crossings. The \timingcut timing requirement is removed to retain more events to give a more accurate determination of the background. A simple scaling can be used to predict the expected cosmic-ray event rate within the timing window because the arrival time of cosmic-ray muons is uniformly distributed. It is found that about 5\% of cosmic-ray events firing the trigger and containing two jets are events where both jets satisfy the \timingcut requirement.

Two additional corrections are applied to determine the final background prediction due to cosmic-ray events. The first accounts for the different live-times of the triggers. The number of empty crossings is 2.9 times smaller than the number used to collect the full data of \intlumi. The second correction weights each event to account for soft tracks due to pile-up and underlying-event effects that would have caused the jet to fail the track isolation requirement had it occurred in a collision environment. To determine the weights a trigger that selects random collision events is used to determine the probability as a function of $\eta$ that a track with $\pt>1$~\GeV is present in a $\Delta R < 0.2$ cone anywhere in the detector as a function of $\eta$.
This probability is applied to each jet in each event to determine an event weight.
The event weights range from 0.55 to 0.63. Combining all the corrections results in a predicted number of cosmic-ray events of $0.3 \pm 0.2$ (statistical error).

Another possible background contribution comes from a beam-halo muon that undergoes bremsstrahlung in the HCal. Two selection criteria reduce this type of background. A jet-timing requirement is imposed because most of the jets produced by beam-halo interactions are not coincident in time with jets from \pp interactions. In addition, events are rejected when track segments in the endcap muon chambers, from the entering beam-halo muon, align in $\phi$ with a jet. These two requirements reduce the background considerably with no discernible effect on the signal.

Unpaired isolated crossings, i.e. crossings where only protons from a single beam are present and at least three buckets on either side of the empty beam's bucket are also empty, can be used to study beam-halo events. To estimate this background, artificial events are created by sampling two jets from a collection of jets passing both a CalRatio trigger active only during unpaired isolated crossings and the leading jet requirements from unpaired isolated crossings. All possible pairs of jets are used and the $\met < 50$ \GeV requirement is applied to each constructed event. The number of jets passing the jet analysis selection and the fraction of constructed events satisfying the \met requirement are combined to estimate the background. This method, which also accounts for cosmic-ray muon contamination, predicts $0.07 \pm 0.07$ events. The large uncertainty is due primarily to the small number of jets passing all required cuts.

Backgrounds from combinations of these non-beam interactions, i.e. a beam-halo jet plus a multi-jet, or a beam-halo jet plus a jet due to a cosmic-ray muon, were found to be negligible.

\section{Systematic uncertainties}
\label{sec:systematics}

Table~\ref{tab:allsys} presents a summary of systematic uncertainties associated with the signal sample. The overall uncertainty, taken as the sum in quadrature of all positive and negative contributions respectively, is listed in the last column. The MC signal samples' statistical uncertainty is shown in Table~\ref{tab:summary} and it is accounted for in the statistical analysis.
The overall normalisation uncertainty of the integrated luminosity is 2.8$\%$ obtained following the same methodology as that detailed in Ref.~\cite{2011lumi} from a preliminary calibration of the luminosity scale derived from beam-separation scans performed in November 2012. The uncertainties on the Higgs boson production cross-sections at \rts~= 8~\tev, which are equal to the uncertainties on the \PhiSM production cross-sections, are about 10\%~\cite{HiggsXs}.

The uncertainty on the signal MC samples due to parton distribution functions (PDF) is calculated by reweighting each event using three different PDF sets (MSTW2008nlo68cl~\cite{mstw}, CT10~\cite{ct10}, and NNPDF2.3~\cite{nnpdf}) and their associated error sets. The RMS change in acceptance for the error sets of each PDF is calculated and combined with the difference in acceptances for each of the three PDFs.

Pile-up primarily affects the acceptance by adding extra tracks and degrading the track isolation of a jet. All MC samples are reweighted to reproduce the observed distribution of the number of interactions per bunch crossing in the data. To determine if pile-up is simulated properly in the MC samples, a direct comparison of data and MC multi-jet samples is performed. The jet \et, EMF, $\eta$, $\phi$, associated tracks and timing distributions as a function of the mean number of pile-up interactions are compared in data and MC simulation. A 10\% systematic uncertainty is assigned to the acceptance covers all the observed differences.

The jet energy scale (JES) uncertainty is evaluated as a function of the jet EMF and $\eta$, following the same strategy used in the {\it in situ} jet energy intercalibration~\cite{JES}. The JES is rederived for low EMF jets for this analysis. The relative jet calorimeter response is studied by balancing the transverse momenta of dijets. The systematic uncertainty is obtained by comparing the \pt-balance in data to the \pt-balance in MC samples. This difference is used to calculate a difference in the JES in data and MC simulation and is propagated to the signal MC samples to get a systematic uncertainty on the acceptance. This study also provides a useful performance comparison between data and MC jets that resemble signal jets.

The \met uncertainty accounts for variations in missing transverse momentum scale and resolution~\cite{met}.  The timing systematic accounts for mismodeling of jet timing between MC and data.  Both of these uncertainties were determined by smearing the associated cut to determine the impact on the acceptance.

The simulation of the trigger is verified by comparing the performance of the trigger in data with the performance in MC simulation on the same multi-jet sample used to evaluate the multi-jet background. Each trigger requirement is studied individually: the jet \et, the \calrat, and the track isolation. Each requirement is adjusted to make the performance match in data and MC events, and the resulting differences in acceptance from the nominal acceptance, for each requirement, are added in quadrature to determine the systematic uncertainty.

The simulation of initial state radiation (ISR) cannot be directly verified because it is difficult to uniquely identify ISR jets in data~\cite{ISR}.
An incorrect ISR rate in the simulation impacts the acceptance by altering the number of jets in the event and by altering the boost of the \PhiSM boson.
Each of these is studied independently.
The ISR jet population is altered event by event so that the number of ISR jets is halved or doubled (jets in MC samples are labelled as containing ISR if they contain a gluon with $\pt > 2$~\GeV).
The population of \vpion jets is not altered by this process, but an added ISR jet may overlap with one of the \vpion jets.
The effect of a boost caused by an ISR jet is studied by exploiting the correlation between the \vpion jet \et and the \PhiSM boost.
From Ref.~\cite{ISR}, the \PhiSM \pt spectrum has an uncertainty of 5\%, which directly correlates with a 5\% uncertainty in the \vpion jet energy.
To calculate the systematic uncertainty associated with the boost, the \pt of ISR jets is conservatively varied by 5\% and the change in acceptance is observed.

The changes in acceptance from both sources of ISR uncertainty are taken as correlated systematic errors and added to get the total systematic for ISR simulation.

An incorrect simulation of final state radiation (FSR) has a negligible effect on the analysis' acceptance. FSR can occur in a prompt or displaced jet. But even if displaced, the extra jet cannot degrade track isolation or deposit extra energy in the ECal if the \vpion has decayed in the HCal.

\begin{table}[ht]
 \caption{Summary of systematic uncertainties for the \PhiSM and Higgs boson production cross-section, jet energy scale, trigger, missing transverse momentum, and the requirement on jet timing as a percentage of the signal yield. Systematic errors that have common values across samples are not listed (pile-up at 10\%, ISR at $^{+2.9}_{-1.2}$\%, and PDF at 2.1\%). The last column reports the total systematic uncertainty (including the luminosity and common systematic errors).}
  \label{tab:allsys}
\begin{center}
 \subfigure{
  \begin{tabular}{l|c|c|c|c|c|c}
    \hline
    Sample                      & H $\sigma$   & JES   & Trigger & \met  & Time & Total\\
    \mHiggs, \mvpion & {[}\%] & [\%] & [\%]    & [\%] & Cut & [\%]\\
    {[}\GeV]                          &                         &          &             &          & [\%] & \\ \hline\hline
    & & & & & \\[-1.5ex]
    126, 10 & $^{+10.4}_{-10.4}$ & $^{+2.2}_{-2.7}$ & $\pm$1.1 & $^{+5.5}_{-2.4}$ & $^{+1.6}_{-6.6}$ & $^{+16.4}_{-16.7}$ \\[1ex]
    126, 25 & $^{+10.4}_{-10.4}$ & $^{+1.5}_{-1.6}$ & $\pm$1.3 & $^{+3.1}_{-1.8}$ & $^{+0.8}_{-3.3}$ & $^{+15.6}_{-15.5}$\\[1ex]
    126, 40 & $^{+10.4}_{-10.4}$ & $^{+2.6}_{-6.2}$ & $\pm$1.1 & $^{+7.7}_{-4.6}$ & $^{+1.9}_{-5.9}$ & $^{+18.2}_{-16.9}$\\[1ex]
\hline\hline
  \end{tabular}
 }
 \subfigure{
  \begin{tabular}{l|c|c|c|c|c|c}
    \hline
    Sample                      & \PhiSM $\sigma$   & JES   & Trigger & \met  & Time & Total\\
    $m_{\PhiSM}$, \mvpion & {[}\%] & [\%] & [\%]    & [\%] & Cut & [\%]\\
    {[}\GeV]                          &                         &          &             &          & [\%] & \\ \hline\hline
    & & & & & \\[-1.5ex]
    100, 10 & $^{+11.1}_{-10.6}$ & $^{+2.3}_{-4.0}$ & $\pm$0.1 & $^{+4.6}_{-3.4}$ & $^{+2.7}_{-9.5}$ & $^{+16.7}_{-18.5}$ \\[1ex]
    100, 25 & $^{+11.1}_{-10.6}$ & $^{+5.5}_{-3.7}$ & $\pm$1.2 & $^{+3.4}_{-2.5}$ & $^{+1.7}_{-0.7}$ & $^{+17.0}_{-15.8}$ \\[1ex] \hline
    & & & & & \\[-1.5ex]
    140, 10 & $^{+10.1}_{-10.3}$ & $^{+0.6}_{-1.1}$ & $\pm$0.5 & $^{+4.0}_{-5.6}$ & $^{+1.9}_{-6.6}$ & $^{+15.6}_{-17.2}$\\[1ex]
    140, 20 & $^{+10.1}_{-10.3}$ & $^{+1.2}_{-1.6}$ & $\pm$1.0 & $^{+4.0}_{-3.9}$ & $^{+0.4}_{-5.0}$ & $^{+15.5}_{-16.2}$\\[1ex]
    140, 40 & $^{+10.1}_{-10.3}$ & $^{+1.3}_{-1.6}$ & $\pm$1.5 & $^{+6.3}_{-4.6}$ & $^{+1.8}_{-2.4}$ & $^{+16.5}_{-15.8}$\\[1ex] \hline
    & & & & & \\[-1.5ex]
    300, 50 & $^{+9.6}_{-10.0}$   & $^{+0.1}_{-0.3}$ & $\pm$0.3 & $^{+9.0}_{-7.4}$ & $^{+0.5}_{-3.0}$ & $^{+13.9}_{-13.3}$ \\[1ex] \hline
    & & & & & \\[-1.5ex]
    600, 50 & $^{+11.2}_{-10.1}$  & $^{+0.0}_{-0.1}$ & $\pm$0.2 & $^{+11.7}_{-11.3}$ & $^{+2.2}_{-4.4}$ & $^{+17.0}_{-16.2}$ \\[1ex]
    600, 150 & $^{+11.2}_{-10.1}$ & $^{+0.2}_{-0.2}$ & $\pm$0.3 & $^{+11.5}_{-10.2}$ & $^{+2.7}_{-5.3}$ & $^{+17.5}_{-15.1}$ \\[1ex] \hline
    & & & & & \\[-1.5ex]
    900, 50 & $^{+12.8}_{-11.5}$  & $^{+0.0}_{-0.1}$ & $\pm$0.1 & $^{+12.6}_{-9.7}$ & $^{+1.0}_{-3.7}$ & $^{+18.5}_{-15.9}$\\[1ex]
    900, 150 & $^{+12.8}_{-11.5}$ & $^{+0.2}_{-0.3}$ & $\pm$0.2 & $^{+11.8}_{-10.9}$ & $^{+0.9}_{-2.5}$ & $^{+18.1}_{-16.3}$\\[1ex]
\hline\hline
  \end{tabular}
 }
\end{center}
\end{table}

\section{Results and exclusion limits}

The global acceptance of the selected event topology in the signal MC samples is a function of $\mPhiSM$, $\mvpion$ and the proper decay length of the \vpion. At a proper decay length of 1.5~m the acceptance ranges from 0.07\% to 0.61\%. The main efficiency loss is due to the low probability that both $\vpion$ decay inside the calorimeter. High mass samples suffer further efficiency loss due to the $\met$ requirement. Table~\ref{tab:summary} lists the expected number of events from all signal MC samples and the background expectation in \intlumi. The $\mHiggs=126$ \GeV mass samples use the SM Higgs boson cross-sections of $\sigmaSM$ = 19.0~pb for the gluon fusion process: other production modes are ignored. The number of events observed in data, 24, is also shown for comparison. No excess of events is observed since the expected background is $23.5\pm 8.0$. The $CL_s$ method~\cite{ALREAD} is used to derive an upper limit on the $\sigma(\Phi) \times$BR$(\Phid)$. A profile likelihood ratio is used as the test statistic and a frequentest calculator is used to generate toy data. The likelihood includes a Poisson probability term describing the total number of observed events. Systematic uncertainties are incorporated as nuisance parameters through their effect on the mean of the Poisson functions and through convolution with their assumed Gaussian distributions. The number of expected events in signal MC samples, together with the estimate of expected background, the observed collision events and all the systematic uncertainties are provided as input for computing the $CL_s$ value, which represents the probability for the given observation to be compatible with the signal + background hypothesis.

\begin{table}[!ht]
\caption{Summary of expected number of signal events, expected background present in the data sample, and the observed number of events in 20.3~fb$^{-1}$. The global acceptance is also given.  The error on the signal samples is statistical only, the error on the expected background is statistical $\oplus$ systematic. All results are normalised for a proper decay length of the \vpion of 1.5~m. A 100\% branching ratio for \PhiHSd is assumed.}
  \label{tab:summary}
  \centering
\subfigure{\small
  \begin{tabular}{l|c|c}
    \hline
   Sample & Expected & Global\\
   (\mHiggs, \mvpion [\GeV]) & yields & acceptance (\%)\\\hline\hline
   126, 10 & 536 $\pm$ 23 &  0.139 $\pm$ 0.006 \\
   126, 25 & 941 $\pm$ 44 &  0.244 $\pm$ 0.011 \\
   126, 40 & 365 $\pm$ 31 &  0.095 $\pm$ 0.008 \\\hline \hline
  \end{tabular}
}
\subfigure{\small
  \begin{tabular}{l|c|c}
    \hline
   Sample & Expected & Global\\
   (\mHiggs, \mvpion [\GeV]) & yields & acceptance (\%)\\\hline\hline
   100, 10 & 440 $\pm$ 29 &  0.073 $\pm$ 0.005  \\
   100, 25 & 424 $\pm$ 37 &  0.070 $\pm$ 0.006  \\\hline
   140, 10 & 525 $\pm$ 20 &  0.168 $\pm$ 0.006 \\
   140, 20 & 900 $\pm$ 37 &  0.287 $\pm$ 0.012 \\
   140, 40 & 641 $\pm$ 30 &  0.205 $\pm$ 0.010 \\\hline
   300, 50 & 444 $\pm$ 11  & 0.609 $\pm$ 0.015\\\hline
   600, 50 & 35 $\pm$ 1 & 0.330 $\pm$ 0.010  \\
   600, 150 & 41 $\pm$ 2 & 0.386 $\pm$ 0.015 \\\hline
   900, 50 & 3.5 $\pm$ 0.1 & 0.304 $\pm$ 0.011 \\
   900, 150 & 4.6 $\pm$ 0.2 & 0.397 $\pm$ 0.016  \\\hline\hline
  \end{tabular}
}
\subfigure{\small
  \begin{tabular}{l|c}
    \hline
    Background & Expected events \\ \hline\hline
    SM Multi-jets & 23.2 $\pm$ 8.0 \\\hline
    Cosmic rays & 0.3 $\pm$ 0.2 \\\hline
    Total Expected Background & $23.5\pm 8.0$ \\\hline\hline
    Data &  24 \\\hline
  \end{tabular}
}
\end{table}

\begin{figure}
  \centering
\subfigure[] {\includegraphics[width=0.48\textwidth]{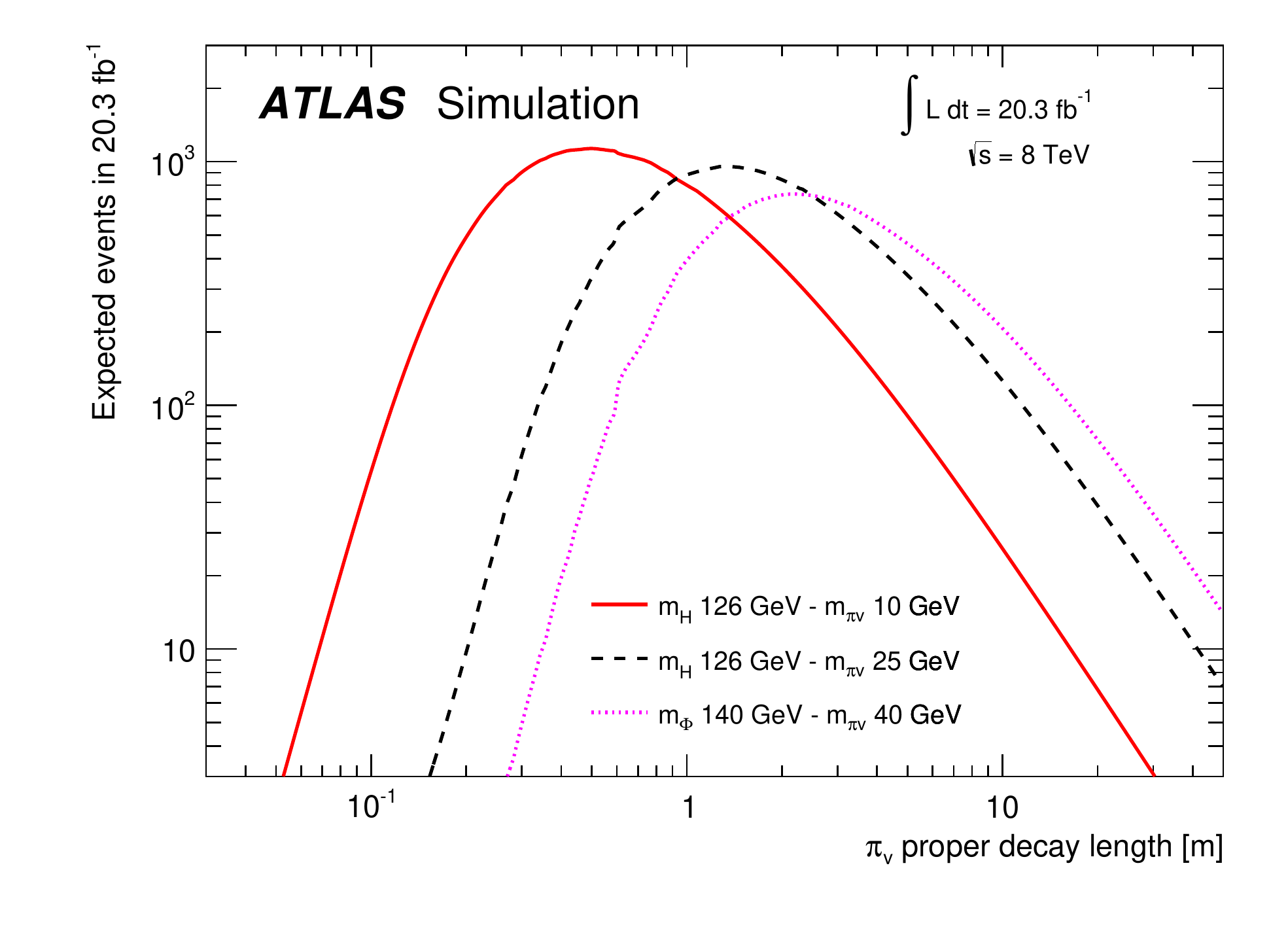}}
\subfigure[] {\includegraphics[width=0.48\textwidth]{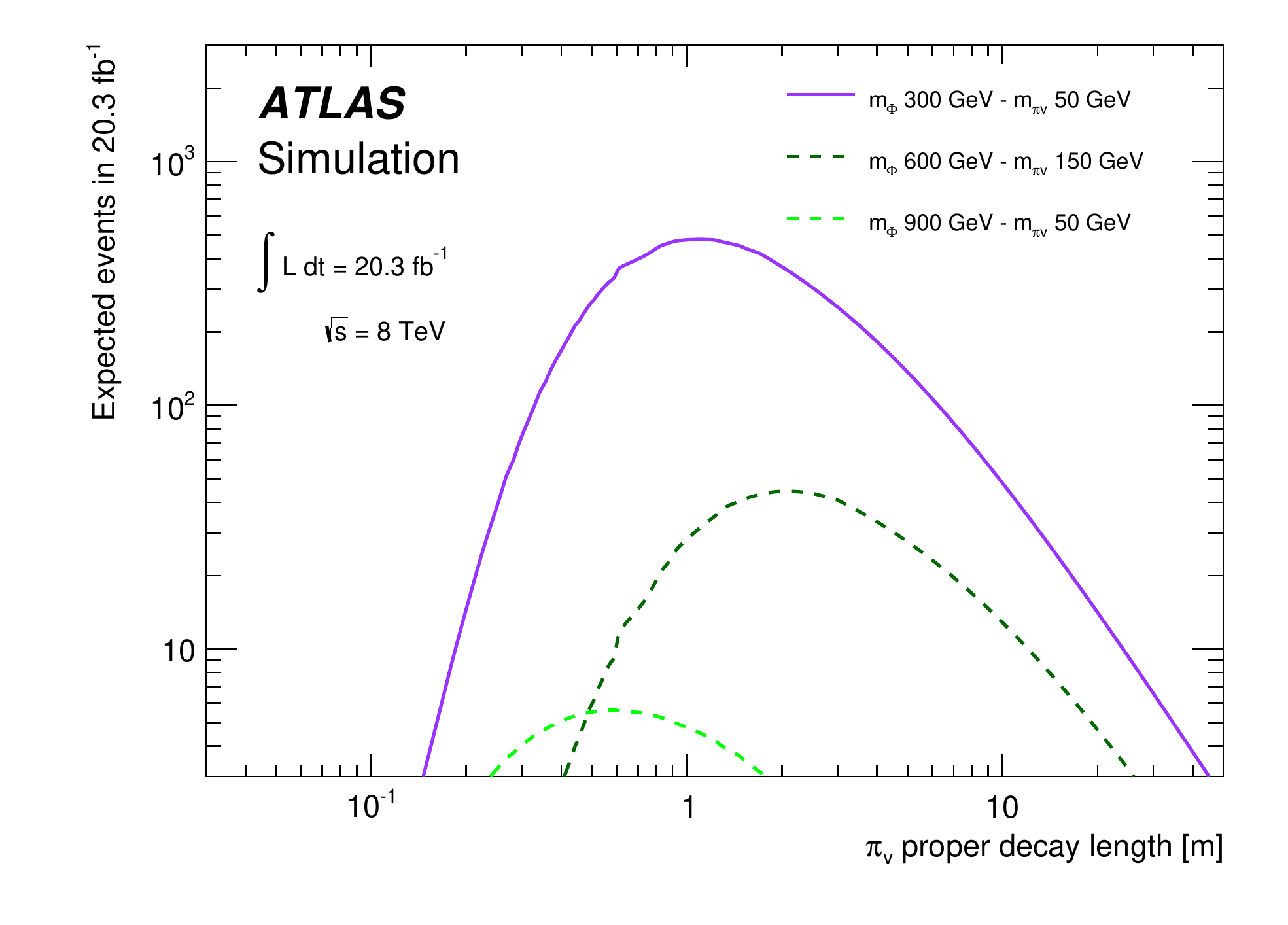}}
\caption{Number of events expected to pass the analysis selection in 20.3~fb$^{-1}$ as a function of the \vpion proper decay length
for (a) three low-mass datasets and (b) three high-mass datasets.
100\% branching ratios
for
\PhiHSd are assumed.}
\label{fig:expEvts}
\end{figure}

\begin{figure}[!ht]
  \centering
\includegraphics[width=0.48\textwidth]{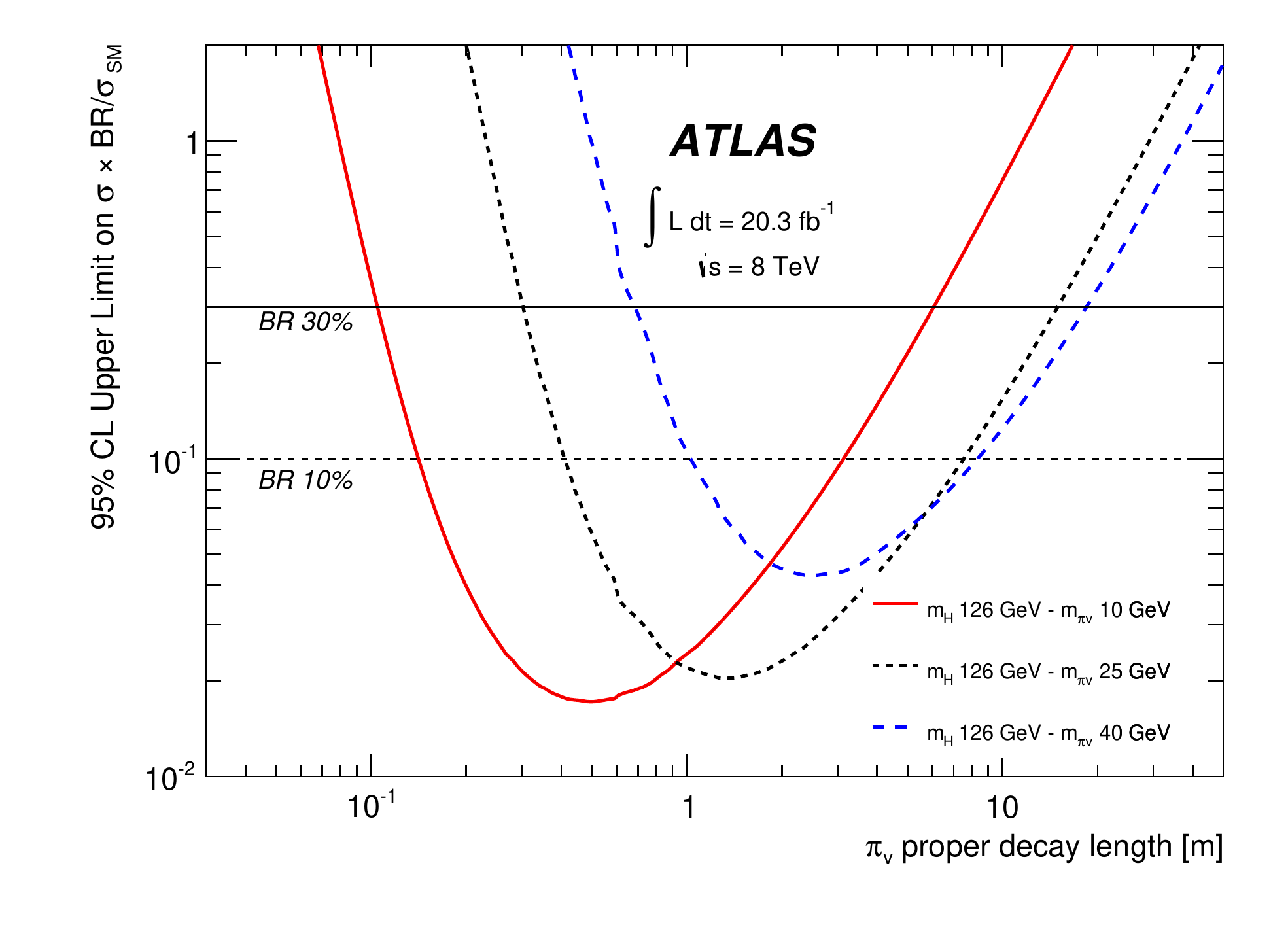}
\caption{Observed 95\% CL limits on $\sigma / \sigmaSM$ for $\mHiggs = 126$~\GeV as a function of the \vpion proper decay length:
the solid line is for $\mvpion = 10$~\GeV, the short-dashed line is for $\mvpion = 25$~\GeV,
the long-dashed line is for $\mvpion = 40$~\GeV.
The $\sigmaSM$ is taken to be $19.0$~pb. The horizontal solid line corresponds to BR = 30\%
and the horizontal dashed line to BR = 10\%.}
\label{fig:obs_126}
\end{figure}

\begin{figure}[!ht]
  \centering
  \subfigure[Observed 95\% CL limits for \mPhiSM = 100 and 140 \GeV.]{\includegraphics[width=0.48\textwidth]{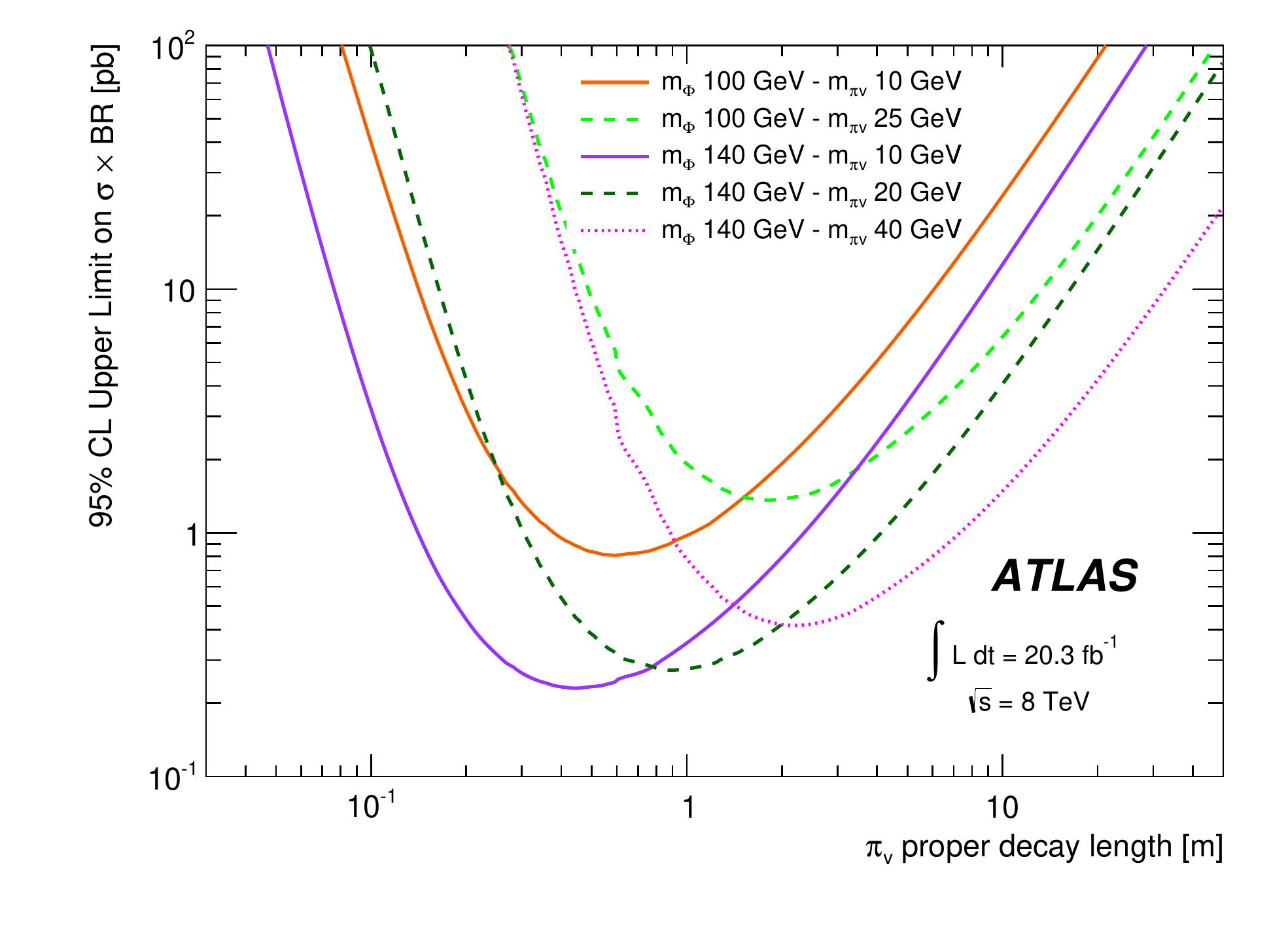}}
  \subfigure[Observed 95\% CL limits for \mPhiSM = 300, 600, and 900 \GeV.]{\includegraphics[width=0.48\textwidth]{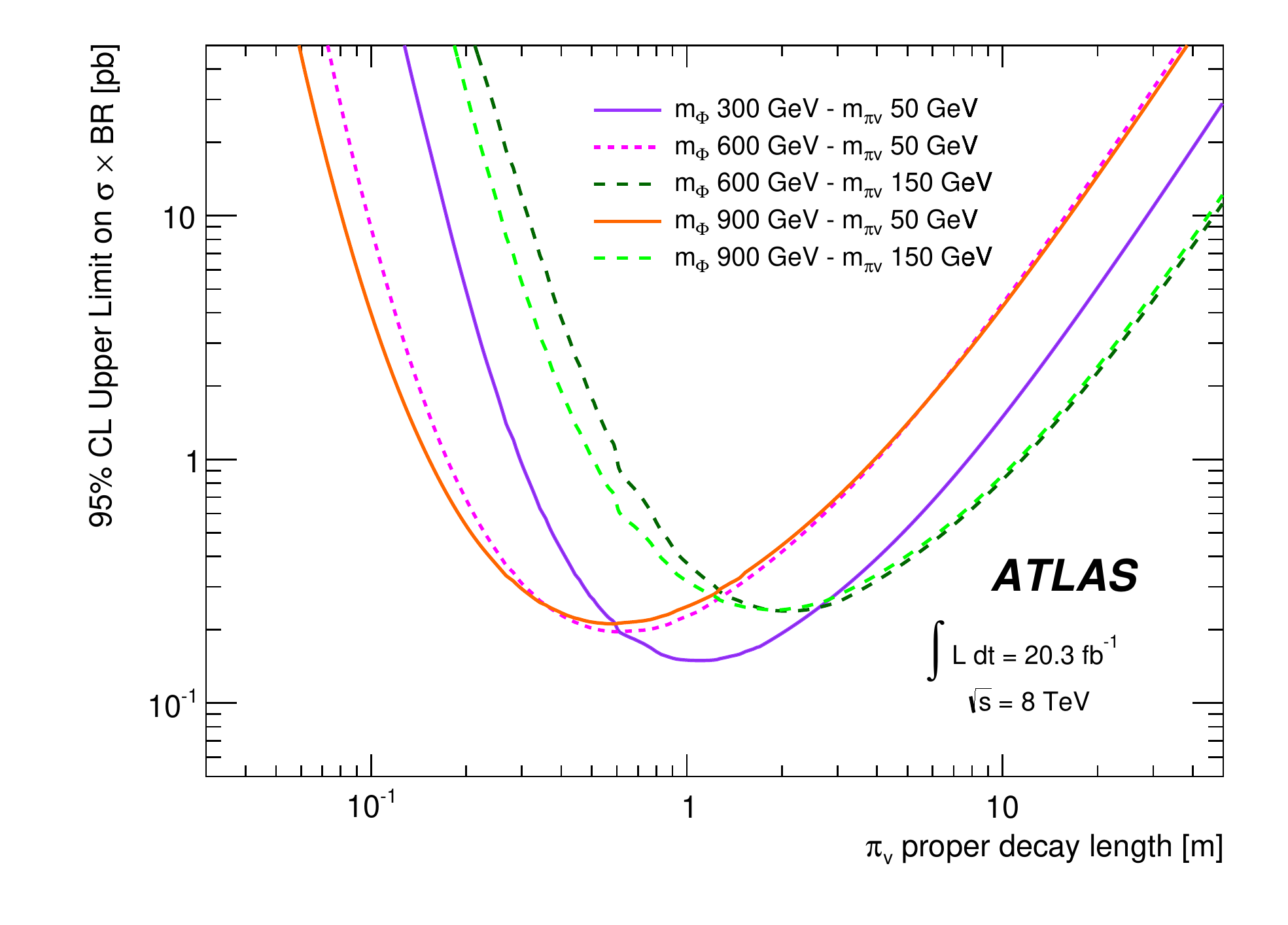}}
\caption{Observed 95\% CL limits on $\sigma \times$ BR [pb] for (a) low and (b) high-mass \PhiSM samples.}
\label{fig:obs_140}
\end{figure}

The acceptance is a function of the \PhiSM mass, the \vpion mass and \vpion~proper decay length. To extrapolate to the number of expected events at different proper decay lengths, a large sample of \vpion~decays is
generated in a range from 0 to 50~m and an efficiency map as a function of \vpion boost is used to determine the efficiency at each decay length.
The resulting efficiencies are then converted into the final number of expected events shown in Fig.~\ref{fig:expEvts}.
Finally, Fig.~\ref{fig:obs_126} and Fig.~\ref{fig:obs_140}
show the observed limit distribution for the three 126~\GeV Higgs samples and for the other \PhiSM samples respectively.
The derived 95\% confidence level (CL) excluded ranges of proper decay length are listed in Table~\ref{tab:exranges} for the $\mHiggs = 126$~\GeV samples,
under the alternative assumptions of a 30\% BR or a 10\% BR for \higgsd.

\begin{table}[ht!]
  \caption{Ranges of \vpion~proper decay lengths excluded at 95\% CL assuming a 30\% and a 10\% BR for a $\mHiggs = 126$ \GeV.}
  \label{tab:exranges}
  \centering
  \begin{tabular}{c|c|c}
    \hline
    MC sample& Excluded range  & Excluded range\\
    \mHiggs, \mvpion & 30\% BR \higgsd & 10\% BR \higgsd\\
    {[}\gev] &[m]  & [m]  \\\hline\hline
    126, 10 & 0.10 -- 6.08 & 0.14 -- 3.13\\\hline
    126, 25 & 0.30 -- 14.99 & 0.41 -- 7.57\\\hline
    126, 40 & 0.68 -- 18.50 & 1.03 -- 8.32\\\hline\hline
  \end{tabular}
\vspace{0.1in}
\end{table}

\FloatBarrier
\section{Summary and conclusions}

A search for the decay of a scalar boson in the mass range from 100~\gev to 900~\gev, including a search for an exotic decay of the Higgs boson, to a pair of long-lived neutral particles decaying in the ATLAS hadronic calorimeter has been presented.
The analysis is based on \intlumi of \pp collisions at \rts~= 8~\tev collected in 2012 by the ATLAS experiment at the LHC.

No significant excess of events is observed over the background estimate.
Limits are set on the \vpion~proper decay lengths for different scalar boson and \vpion~mass combinations. For a SM Higgs decaying to \vpion proper decay lengths between 0.10 m and 18.50 m assuming a 30\% BR are ruled out, and between 0.14 m and 8.32 m assuming a BR of 10\%. Results for low mass \PhiSM (100 \gev and 140 \gev) and high mass \PhiSM (300 \gev, 600 \gev, and 900 \gev) have also been presented as a function of proper decay length.

\section*{Acknowledgements}
We thank CERN for the very successful operation of the LHC, as well as the
support staff from our institutions without whom ATLAS could not be
operated efficiently.

\indent We acknowledge the support of ANPCyT, Argentina; YerPhI, Armenia; ARC,
Australia; BMWFW and FWF, Austria; ANAS, Azerbaijan; SSTC, Belarus; CNPq and FAPESP,
Brazil; NSERC, NRC and CFI, Canada; CERN; CONICYT, Chile; CAS, MOST and NSFC,
China; COLCIENCIAS, Colombia; MSMT CR, MPO CR and VSC CR, Czech Republic;
DNRF, DNSRC and Lundbeck Foundation, Denmark; EPLANET, ERC and NSRF, European Union;
IN2P3-CNRS, CEA-DSM/IRFU, France; GNSF, Georgia; BMBF, DFG, HGF, MPG and AvH
Foundation, Germany; GSRT and NSRF, Greece; RGC, Hong Kong SAR, China; ISF, MINERVA, GIF, I-CORE and Benoziyo Center, Israel; INFN, Italy; MEXT and JSPS, Japan; CNRST, Morocco; FOM and NWO,
Netherlands; BRF and RCN, Norway; MNiSW and NCN, Poland; GRICES and FCT, Portugal; MNE/IFA, Romania; MES of Russia and ROSATOM, Russian Federation; JINR; MSTD,
Serbia; MSSR, Slovakia; ARRS and MIZ\v{S}, Slovenia; DST/NRF, South Africa;
MINECO, Spain; SRC and Wallenberg Foundation, Sweden; SER, SNSF and Cantons of
Bern and Geneva, Switzerland; NSC, Taiwan; TAEK, Turkey; STFC, the Royal
Society and Leverhulme Trust, United Kingdom; DOE and NSF, United States of
America.

\indent The crucial computing support from all WLCG partners is acknowledged
gratefully, in particular from CERN and the ATLAS Tier-1 facilities at
TRIUMF (Canada), NDGF (Denmark, Norway, Sweden), CC-IN2P3 (France),
KIT/GridKA (Germany), INFN-CNAF (Italy), NL-T1 (Netherlands), PIC (Spain),
ASGC (Taiwan), RAL (UK) and BNL (USA) and in the Tier-2 facilities
worldwide.

\bibliographystyle{atlasstylem}
\bibliography{atlaspaper}

\onecolumn
\clearpage
\begin{flushleft}
{\Large The ATLAS Collaboration}

\bigskip

G.~Aad$^{\rm 84}$,
B.~Abbott$^{\rm 112}$,
J.~Abdallah$^{\rm 152}$,
S.~Abdel~Khalek$^{\rm 116}$,
O.~Abdinov$^{\rm 11}$,
R.~Aben$^{\rm 106}$,
B.~Abi$^{\rm 113}$,
M.~Abolins$^{\rm 89}$,
O.S.~AbouZeid$^{\rm 159}$,
H.~Abramowicz$^{\rm 154}$
% ...
\bigskip
\\
% ...
$^{11}$ Institute of Physics, Azerbaijan Academy of Sciences, Baku, Azerbaijan\\
% ...
$^{84}$ CPPM, Aix-Marseille Universit{\'e} and CNRS/IN2P3, Marseille, France\\
% ...
$^{89}$ Department of Physics and Astronomy, Michigan State University, East Lansing MI, United States of America\\
% ...
$^{106}$ Nikhef National Institute for Subatomic Physics and University of Amsterdam, Amsterdam, Netherlands\\
% ...
$^{112}$ Homer L. Dodge Department of Physics and Astronomy, University of Oklahoma, Norman OK, United States of America\\
$^{113}$ Department of Physics, Oklahoma State University, Stillwater OK, United States of America\\
% ...
$^{116}$ LAL, Universit{\'e} Paris-Sud and CNRS/IN2P3, Orsay, France\\
% ...
$^{152}$ Institute of Physics, Academia Sinica, Taipei, Taiwan\\
% ...
$^{154}$ Raymond and Beverly Sackler School of Physics and Astronomy, Tel Aviv University, Tel Aviv, Israel\\
% ...
$^{159}$ Department of Physics, University of Toronto, Toronto ON, Canada\\
% ...
\end{flushleft}

\end{document}